\documentclass[aps,pre,twocolumn,showpacs,superscriptaddress]{revtex4}
\usepackage{amssymb}
\usepackage{bbm}
\usepackage{graphicx}
\usepackage{amsmath,amssymb}

\begin{document}

\preprint{2016 Conference paper}

\title{Order Parameter Analysis of Synchronization transitions on star networks}
\author{Hongbin Chen}
\affiliation{Institute of Systems Science, Huaqiao University, Xiamen 361021, China}
\affiliation{College of Information Science and Engineering, Huaqiao University, Xiamen 361021, China}
\author{Yuting Sun}
\affiliation{Department of Physics and the Beijing-Hong Kong-Singapore Joint Centre for Nonlinear and Complex Systems (Beijing), Beijing Normal University, Beijing 100875, China}
\author{Jian Gao}
\affiliation{Department of Physics and the Beijing-Hong Kong-Singapore Joint Centre for Nonlinear and Complex Systems (Beijing), Beijing Normal University, Beijing 100875, China}
\author{Can Xu}
\email{xushecan@163.com}
\affiliation{Department of Physics and the Beijing-Hong Kong-Singapore Joint Centre for Nonlinear and Complex Systems (Beijing), Beijing Normal University, Beijing 100875, China}
\author{Zhigang Zheng}
\email{zgzheng@hqu.edu.cn}
\affiliation{Institute of Systems Science, Huaqiao University, Xiamen 361021, China}
\affiliation{College of Information Science and Engineering, Huaqiao University, Xiamen 361021, China}
\date{\today}

\begin{abstract}
Collective behaviors of populations of coupled oscillators have attracted much attention in recent years. In this paper, an order parameter approach is proposed to study the low-dimensional dynamical mechanism of collective synchronizations by adopting the star-topology of coupled oscillators as a prototype system. The order parameter equation of star-linked phase oscillators can be obtained in terms of the Watanabe-Strogatz transformation, Ott-Antonsen ansatz, and the ensemble order parameter approach. Different solutions of the order parameter equation correspond to diverse collective states, and different bifurcations reveal various transitions among these collective states.  The properties of various transitions are revealed in the star-network model by using tools of nonlinear dynamics such as time reversibility analysis and linear stability analysis.
\end{abstract}


\maketitle
\section{Introduction}
\label{intro}
Understanding the intrinsic microscopic mechanism embedded in collective macroscopic behaviors of populations of coupled units on heterogenous networks has become a focus in a variety of fields, such as the biological neurons circadian rhythm, chemical reacting cells, and even society systems~\cite{kuramoto1984chemical,acebron2005kuramoto,strogatz2000kuramoto,pikovsky2002synchronization,dorogovtsev2008critical,arenas2008synchronization,zheng1998phase,zheng2000pre}. Numerous different emerging macroscopic states/phases have been revealed, and various non-equilibrium transitions among these states have been observed and studied on heterogenous networks~\cite{Paley2005,Silber1993,Strogatz1993,lu2014,gomez2011explosive,omel2012nonuniversal,topaj2002reversibility,zhou2015,zhang2013explosive,zheng2004book,yaonan-review}.

The transitions among different collective states on heterogeneous networks exhibit the typical feature of multistability, i. e., these states may coexist for a group of given parameters and depend on the choice of initial conditions. This interesting behavior is closely related to the first-order phase transition, and multistability in the discontinuous transitions indicate the competitions of miscellaneous attractors and their corresponding basins of attraction in phase space. For a network of coupled oscillators, the microscopic description of the dynamics of oscillators should be made in a high-dimensional phase space, which is very difficult to deal with. The key point in understanding macroscopic transitions is the projection of the dynamics from this high-dimensional space to a much lower-dimensional subspace. This can be executed by introducing appropriate order parameters and building their dynamical equations. Ott and Antonsen ~\cite{ott2008low} proposed an ansatz to project the infinite-dimensional dynamics to a low-dimensional manifold called the Ott-Antonsen (OA) manifold, which has been successfully applied to systems composed of large numbers of oscillators. However, strictly speaking, the OA manifold analysis cannot be applied to finite-oscillator systems. Watanabe and Strogatz introduced the M\"{o}bius transformation for finite-size systems with specific symmetries to obtain an exact three-dimensional dynamics ~\cite{ws1993,ws1994}, but this scheme cannot be extended to general finite systems.  The mechanism of the validity of the OA approach  was recently studied, and the ensemble order parameter approach is proposed, which extends the OA approach to more general cases such as a finite-number of oscillators and more general coupling forms ~\cite {gaojian2016eop}.

Abrupt or explosive transition from incoherent state to synchronization may occur on networks if the frequencies of oscillators on nodes are positively correlated to the node's degrees ~\cite{gomez2011explosive}, which has been observed numerically on scale-free networks and experimentally in electronic circuits~\cite{hu2014exact,leyva2012explosive}. The first-order transition can be changed and more ways of transitions can be observed by adjusting the phase shift among oscillators~\cite{xucan2015star}. Numerous efforts have been made to understand the mechanism of explosive synchronization from different viewpoints such as the topological structures of networks, the coupling functions among nodes, and so on~\cite{zhang2013explosive,leyva2012explosive,li2013reexamination,peron2012explosive,ji2013cluster,zhang2014,leyva2013explosive,fop1}. 

It is valuable to analytically understand the transitions among various synchrony states on heterogeneous networks. The star topology is the simplest while the key topology in describing the heterogeneity property of complex networks such as the scale-free networks ~\cite{Bergner2013remote,Oleksandr2012Bifurcations,Theesar2013,Vladimir2015}.  In this paper, we study the collective states and the abundant transitions among these states on a star network by considering the effect of the phase shift among coupled oscillators ~\cite{xucan2015star,xucan2016a,gaojian2016eop,xucan2016b,huangxia2016}. The dynamics of star networks of oscillators is analytically studied by building the equations of motion of the order parameter for networks with a finite size, which accomplishes a great reduction from microscopic high-dimensional phase dynamics of coupled oscillators to a macroscopic low-dimensional dynamics. Based on the order parameter dynamics, we further reveal numerous transitions among different collective states in this model by using tools of nonlinear dynamics such as time reversibility analysis~\cite{topaj2002reversibility} and linear stability analysis. We found three typical processes of the transitions to the synchronous state, i. e., the transitions from the neutral state, the in-phase state or the splay state to the synchronous state, and a continuous process of desynchronization and a group of hybrid phase transitions that are discontinuous with no hysteresis. 

\section{The Ott-Antonsen ansatz and the Watanabe-Strogatz approach}
\label{sec:1}
We first illustrate the Ott-Antonsen ansatz~\cite{ott2008low} briefly by analyzing the following class of identical oscillators governed by the equations of motion
\begin{equation}\label{equ:01}
\dot\varphi_{j}=fe^{i\varphi_{j}}+g+\bar{f}e^{-i\varphi_{j}},\quad j=1,\cdots,N,
\end{equation}
where $f$ is a smooth, complex-valued 2$\pi$-periodic function of the phases $\varphi_{1},\cdots,\varphi_{N}$ and the overbar denotes complex conjugate, $\it{g}$ is a real valued function since $\dot\varphi_{j}$ is real. In the limit $N\longrightarrow\infty$, by introducing the distribution of phases of oscillators, the evolution of the system (1) is given by the continuity equation
 \begin{equation}\label{equ:02}
\frac{\partial{\rho}}{\partial{t}}+\frac{\partial{(\rho\nu)}}{\partial{\phi}}=0,
\end{equation}
where $\rho(\phi,t)$ is the phase distribution function, and $\rho(\phi,t)d\phi$ gives the fraction of phases that lie between $\phi$ and $\phi+d{\phi}$ at time $\it{t}$. The velocity field is the Eulerian version of Eq. (\ref{equ:01}),
\begin{equation}\label{equ:03}
\nu(\phi,t)=fe^{i\varphi}+g+\bar{f}e^{-i\varphi}.
\end{equation}

 Suppose $\rho$ is of the form
 \begin{equation}\label{equ:04}
\rho(\phi,t)=\frac{1}{2\pi}\{1+\sum_{n=1}^{\infty}({\bar{z}(t)^n}{e^{in\phi}}+z(t)^n{e^{-in\phi}})\}
\end{equation}
 for some unknown function $z$ that is independent of $\phi$. Note that Eq. (\ref{equ:04}) is just an algebraic rearrangement of the usual form for the Poisson kernel
 \begin{equation}\label{equ:05}
\rho(\phi)=\frac{1}{2\pi}\frac{1-r^2}{1-2r\cos(\phi-\Phi)+r^2},
\end{equation}
where the complex number $z$ can be expressed in the complex plane as
 \begin{equation}\label{equ:06}
z=r{e^{i\Phi}}.
\end{equation}
The ansatz (\ref{equ:04}) defines a submanifold in the infinite-dimensional space of the density function $\rho$. This Poisson submanifold is two-dimensional and is parameterized by the complex number $z$, or equivalently, by the polar coordinates $r$ and $\Phi$. An intriguing point discovered by Ott et al. ~\cite{ott2008low} is the invariance of the Poisson submanifold, i. e., if the initial phase density is a Poisson kernel, it remains a Poisson kernel for all the time. This can be verfied by substituting the velocity field (\ref{equ:03}) and the ansatz (\ref{equ:04}) into the continuity equation (\ref{equ:02}). It can be found that the amplitude equations for each harmonic $e^{in\phi}$ are simultaneously satisfied if and only if $z(t)$ evolves according to
\begin{equation}\label{equ:07}
\dot{z}=i(f{z}^2+gz+\bar{f}).
\end{equation}
This equation can be recast in a more physically meaningful form in terms of the complex order parameter defined as the centroid of the phases $\phi$ regarded as points $e^{i\phi}$ on the unit circle:
\begin{equation}\label{equ:08}
<e^{i\phi}>=\int_0^{2\pi}{e^{i\phi}\rho(\phi,t)}d{\phi}.
\end{equation}
By substituting Eq. (\ref{equ:04}) into Eq. (\ref{equ:08}) one may find that
\begin{equation}\label{equ:09}
 z=<e^{i\phi}>=r{e^{i\Phi}}
\end{equation}
 for all states on the Poisson submanifold. Then the meaning of $z$ is clear that it represents the order parameter of the system, $r$ is the modulus of it and $\Phi$ is the mean phase of it. However, whether the governing equation Eq. (\ref{equ:07}) can be used for system with finite size can not be implied from Ott-Antonsen ansatz.

 For a finite  number of oscillators $N$, the original microscopic dynamical state can be reduced to a macroscopic collective state by the Watanabe-Strogatz approach~\cite{ws1993,ws1994}, and the governing equations of the system can also be generated by  the M\"{o}bius group action~\cite{charles1995,marvel2009identical}. The class of identical oscillators still governed by the equations of motion Eq. (\ref{equ:01}), then the oscillators' phases $\varphi_{j}(t)$ evolve according to the action of the M\"{o}bius group on the complex unit cycle
 \begin{equation}\label{equ:10}
 e^{i{\varphi_{j}(t)}}=M_{t}{(e^{i\theta_j})}
\end{equation}
 for $j=1,\ldots,N$, where $M_t$ is a one-parameter family of M\"{o}bius transformations and $\theta_j$ is a constant angle. By parameterizing the one-parameter family of M\"{o}bius transformations as
 \begin{equation}\label{equ:11}
 M_{t}(w)=\frac{e^{i\psi}w+\eta}{1+\bar{\eta}e^{i\psi}w},
\end{equation}
where $|\eta(t)|<1$ and $\psi(t)\in{R}$, and let
 \begin{equation}\label{equ:12}
 w_j=e^{i\theta_j}.
\end{equation}
One then obtains
\begin{subequations} \label{equ:18}
\begin{align}
 \dot{\eta} &=i(f{\eta}^2+g\eta+\bar{f}), \label{equ:18a}\\
 \dot{\psi} &=f{\eta}+g+\bar{f}\bar{\eta}. \label{equ:18b}
 \end{align}
\end{subequations}
With these new variables, one could rewrite the order parameter as
 \begin{equation}\label{equ:19}
 z(t)=\frac{1}{N}\sum_{j=1}^{N}\frac{e^{i\psi}e^{i\theta_j}+\eta(t)}{1+\bar{\eta}(t)e^{i\psi}e^{i\theta_j}},
\end{equation}

Eqs. (\ref{equ:18}) and (\ref{equ:19}) can describe the system with arbitrary initial conditions as $\eta(0),\psi(0)$ and N constants $\theta_j,1\leq j\leq N$. The order parameter (\ref{equ:19}) could be simplified further by choosing the constants
  \begin{equation}\label{equ:20}
 \theta_j=2\pi\frac{j-1}{N},1\leq j\leq N,
\end{equation}
with which, the order parameter (\ref{equ:19}) reads
  \begin{equation}\label{equ:21}
 z(t)=\eta(t)(1+I),
\end{equation}
 where $I=(1-|\eta(t)|^{-2})/(1\pm(e^{i\psi}\bar{\eta}(t))^{-N})$, "$-$" for the case with even N and "$+$" for the case with odd N. One can verify that for large N, $I\ll1$, the order parameter could be approximated as
 \begin{equation}\label{equ:22}
 z(t)\approx\eta(t), N\gg1.
\end{equation}
Based on the analysis above, the dynamics of the system with finite size can be described by the same equation as  the governing equation (\ref{equ:07}) which obtained from the Ott-Antonsen ansatz for the system with infinite size. Then Eq. (\ref{equ:07}) can still be used to explore the low-dimensional collective behaviors of the system with finite size.

\section{The Sakaguchi-Kuramoto model on star networks: The order parameter equation}
\label{sec:2}
We start with a star network of coupled phase oscillators with nonzero phase shift as our working model. In the star network with one hub and $K$ leaves, the degree of the leaves is $k_i=1$ ($i=1,...,K$) and the degree of the hub is $k_h=K$. Suppose that the natural frequencies of the oscillators are proportional to their degrees, the equations of motion for the hub and leaf nodes read
\begin{equation}\label{equ:23}
\begin{aligned}
\dot{\theta}_{h}&=\omega_{h}+\lambda\, \sum_{j=1}^{K}\sin\,(\theta_{j}-\theta_{h}-\alpha),\\
\dot{\theta}_{j}&=\omega+\lambda\,\sin\,(\theta_{h}-\theta_{j}-\alpha), 1\leq j\leq K,
\end{aligned}
\end{equation}
where $\theta_{h},\theta_{j}$ and $\omega_{h},\omega$ are instantaneous phases and natural frequencies of the hub and leaf nodes respectively. $\lambda$ is the coupling strength. $K$ is the number of leaf nodes connected with this hub and $\alpha$ is the phase shift. The effect of phase shift among coupled oscillators has been extensively investigated in recent years, while this has been seldom discussed in star networks ~\cite{sakaguchi1986soluble,omel2012nonuniversal}. Abundant collective dynamics appear in the global coupled model in the presence of a finite phase shift~\cite{omel2012nonuniversal}, where synchrony can decay or incoherence can regain its stability with increasing coupling and multistability between partially synchronized and/or the incoherent state can appear in the globally coupled network.

The coupled phase oscillator system with $\alpha=0$ of the star network was originally to study the characteristics of the explosive synchronization~\cite{gomez2011explosive}, however the process of the synchronization may be influenced with the introduction of phase shift~\cite{omel2012nonuniversal}. By introducing the phase differences between the hub and leaves $\varphi_{j}=\theta_{h}-\theta_{j}$, the phase dynamics on star networks can be transformed to the following phase difference dynamics on an all-connected network,
\begin{equation}\label{equ:26}
\dot\varphi_{i}=\Delta\omega-\lambda\,\sum_{j=1}^{K}\,\sin(\varphi_{j}+\alpha)-\lambda\,\sin(\varphi_{i}-\alpha),
\end{equation}
where $1\leq i\leq K$. We further define the order parameter of the all-connected network to describe the degree of synchronization as
\begin{equation}\label{equ:27}
z(t)\equiv r(t)e^{i\Phi(t)}=\frac{1}{K}\sum_{j=1}^{K}e^{i(\varphi_{j})}.
\end{equation}
It is worth noting that the star network becomes globally synchronous if the modulus $r(t)=1$ and the mean phase $\Phi(t)=const$. If the modulus $r(t)=1$ while the mean phase $\Phi(t)$ is periodic which corresponds to the state with $\varphi_{j}(t)=\varphi(t)$, all the leaf nodes are synchronous to each other while they are asynchronous to the hub oscillator.

It is instructive to rewrite Eq. (\ref{equ:26}) as
\begin{equation}\label{equ:28}
\dot\varphi_{j}=fe^{i\varphi_{j}}+g+\bar{f}e^{-i\varphi_{j}},\quad j=1,\cdots,K,
\end{equation}
where $i$ denotes the imaginary unit and $f=i\dfrac{\lambda}{2}e^{-i\alpha}$, $g=\Delta\omega-\lambda Kr\sin(\Phi+\alpha)$.

For finite $K$, due to the high topological symmetry of the star network, the collective behaviors of the system can be analyzed by deriving the  low-dimensional dynamical equations in terms of both the ensemble order parameter approach ~\cite{gaojian2016eop} and the the Watanabe-Strogatz transformation~\cite{ws1993,ws1994}. If the initial phases of oscillators are chosen as (\ref{equ:20}), we have the dynamical equation for the order parameter $z(t)$ as
\begin{equation}\label{equ:30}
\dot z=-\dfrac{\lambda}{2}e^{-i\alpha}z^{2}+i(\Delta \omega-\lambda Kr\sin(\Phi+\alpha))z+\dfrac{\lambda}{2}e^{i\alpha},
\end{equation}
which is just the OA result (\ref{equ:23})  in terms of the order parameter. Different solutions of Eq. (\ref{equ:30}) build correspondences with diverse collective states of the coupled oscillator system.

\section{Collective Dynamics of Stationary States}
\label{sec:3}
In the following we start discussing the collective dynamics of the star network in terms of the Eq. (\ref{equ:30}). By setting $z=x+iy$, we can describe the order parameter dynamics in the $x-y$ plane as
\begin{equation}\label{equ:31}
\begin{aligned}
\dot x&=\lambda(\dfrac{1}{2}+K)\cos\alpha\, y^{2}-\dfrac{\lambda}{2}\cos\alpha\, x^{2}\\
      &+\lambda(K-1)\sin\alpha\, x y-\Delta\omega\, y+\dfrac{\lambda}{2}\cos\alpha,\\
\dot y&=\lambda(\dfrac{1}{2}-K)\sin\alpha\, x^{2}-\dfrac{\lambda}{2}\sin\alpha\, y^{2}\\
      &-\lambda(K+1)\cos\alpha\, x y+\Delta\omega\, x+\dfrac{\lambda}{2}\sin\alpha.
\end{aligned}
\end{equation}
 The steady-state solutions are determined by setting $\dot x=0$ and $\dot y=0$, which results in four fixed points noted by $(x_{i},y_{i})$ with
\begin{equation}\label{equ:32}
\begin{aligned}
&x_{1,2}=\frac{-\sin\alpha\Delta\omega\pm A\sin\alpha}{\lambda(2K\cos2\alpha+1)},\\
&y_{1,2}=-\frac{-\cos\alpha\Delta\omega\pm A\cos\alpha}{\lambda(2K\cos2\alpha+1)},   \\
&x_{3,4}=\frac{\sin\alpha}{\lambda}+\frac{\frac{\sin2\alpha}{2}B\pm K(\frac{\sin2\alpha}{2}B-\sin2\alpha^{2})}{\lambda\sin\alpha(K^{2}+2\cos(2\alpha)K+1)},\\
&y_{3,4}=\frac{-\Delta\omega(-\cos\alpha\pm\sin\alpha B-K\cos\alpha)}{\lambda(K^{2}+2\cos(2\alpha)K+1)},
\end{aligned}
\end{equation}
where $"+"$ represents the fixed points $(x_{1,3},y_{1,3})$, $"-"$ represents the fixed points $(x_{2,4},y_{2,4})$ and
\begin{equation}\label{equ:33}
\begin{aligned}
&A=\sqrt{-2K\lambda^{2}\cos2\alpha-\lambda^{2}+\Delta\omega^{2}},\\
&B=\sqrt{\lambda^{2}+K^{2}\lambda^{2}+2K\lambda^{2}\cos2\alpha-\Delta\omega^{2}}.
\end{aligned}
\end{equation}
The existence condition for the fixed points are determined by Eq. (\ref{equ:33}), where $-2K\lambda^{2}\cos2\alpha-\lambda^{2}+\Delta\omega^{2}\geq0$, and $\lambda^{2}+K^{2}\lambda^{2}+2K\lambda^{2}\cos2\alpha-\Delta\omega^{2}\geq0$. For the fixed points $(x_{1,2},y_{1,2})$, the existence condition can be given as
\begin{equation}\label{equ:34}
\lambda\leq\lambda_{1}=\frac{\Delta\omega}{\sqrt{2K\cos2\alpha+1}},
\end{equation}
and for the fixed points $(x_{3,4},y_{3,4})$, the existence condition is
\begin{equation}\label{equ:35}
\lambda\geq\lambda_{2}=\frac{\Delta\omega}{\sqrt{K^{2}+2K\cos2\alpha+1}}.
\end{equation}
For the fixed points $(x_{3,4},y_{3,4})$, there is an additional natural restriction relation
\begin{equation}\label{equ:36}
 x^{2}+y^{2}=1,
\end{equation}
 while for the fixed points $(x_{1,2},y_{1,2})$, $x^{2}+y^{2}$ may be greater or lower than 1. The definition of $z$ implies that only those fixed points satisfying $x^{2}+y^{2}\leq1$ are reasonable.

\begin{figure}
  \includegraphics[width=3in]{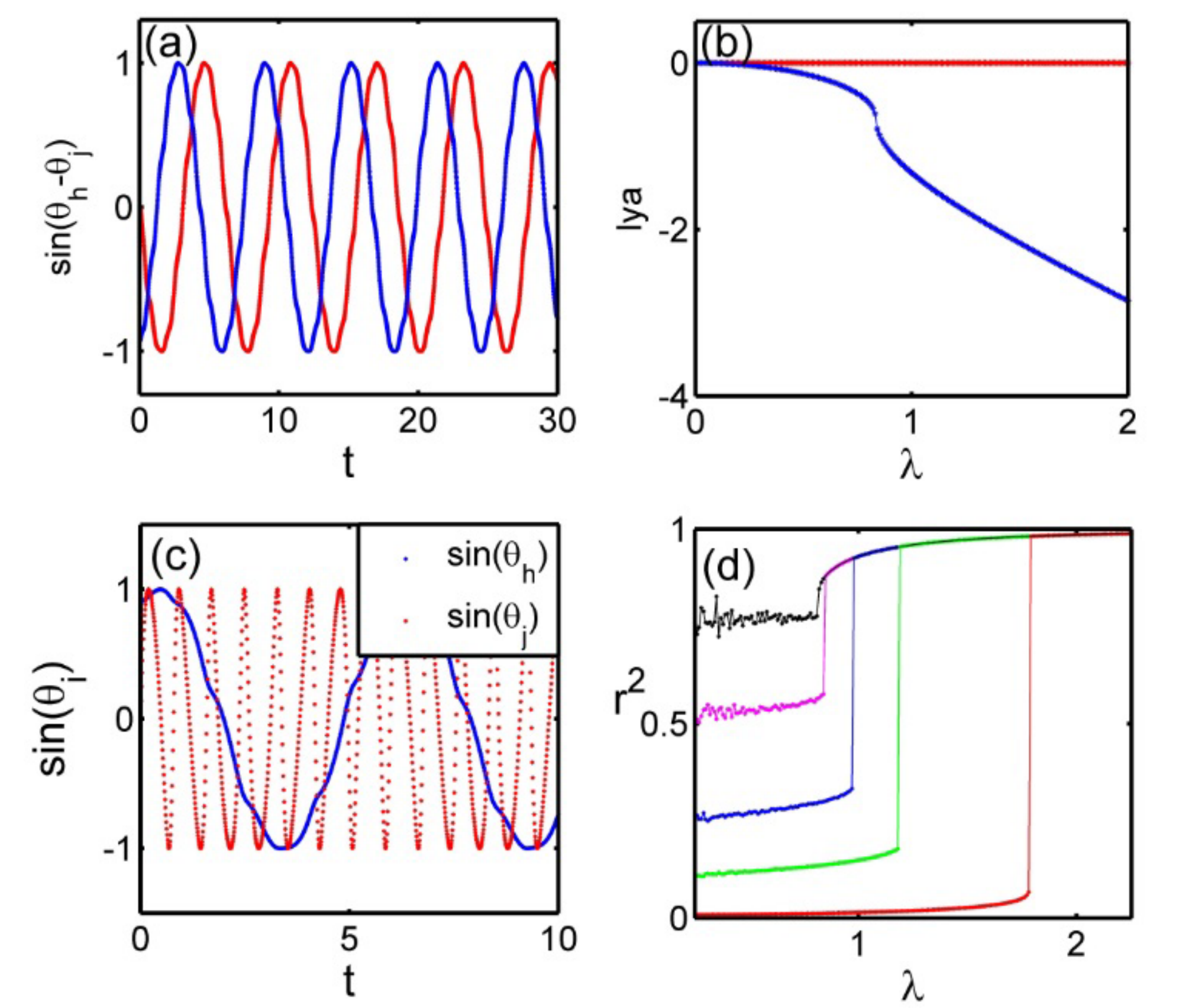}
\caption{(a) The time evolution of sin$\varphi_{j}$(t) with $\alpha=-0.4\pi, \lambda=2$, $j=1,2$. (b) The Lyapunov exponents of the network with $\alpha=0.1\pi$.  (c) The time evolution of sin$\theta_{i}$(t) with $\alpha=0.1\pi, \lambda=0.5, i=1,\cdots,K$. (d) The order parameter against the coupling strength with different initial states for $\alpha=0$. The size of the star network is $N=11$.}
\label{fig:1}       
\end{figure}

Linear stability analysis can be applied to the fixed points $(x_i,y_i),i=1,2,3,4$ by computing the eigenvalues of the $2\times2$ Jacobian matrix $J$ of the fixed points with elements
\begin{equation}\label{equ:37}
\begin{aligned}
&J_{11}=-\lambda\cos\alpha x_i+\lambda(K-1)\sin\alpha y_i,\\
&J_{12}=\lambda(1+2K)\cos\alpha y_i+\lambda(K-1)\sin\alpha x_i-\Delta \omega,\\
&J_{21}=\lambda(1-2K)\sin\alpha x_i-\lambda(K+1)\cos\alpha y_i+\Delta\omega,\\
&J_{22}=-\lambda\sin\alpha y_i-\lambda(K+1)\cos\alpha x_i.
\end{aligned}
\end{equation}
The eigenvalues of the Jacobian matrix are
\begin{equation}\label{equ:38}
\beta_{1,2}=\frac{J_{11}+J_{22}\pm\sqrt{(J_{11}+J_{22})^{2}-4(J_{11}J_{22}-J_{12}J_{21})}}{2}.
\end{equation}
The stability conditions of the four fixed points are summarized in Table \ref{tabel:1}, where The parameters in the table are $\hat\lambda_{c}^{f}=\footnotesize{{\Delta\omega}/{\sqrt{2K\cos2\alpha+1}}}$,
 $\lambda_{sc}^{+}=\footnotesize{{-\Delta\omega}/({K\cos2\alpha+1}})$, $\lambda_{sc}^{-}=\footnotesize{{\Delta\omega}/({K\cos2\alpha+1}})$,
 $\alpha_{0}^{-}=-\arccos(-1/K)/2$, $\alpha_{0}^{+}=\arccos(-1/K)/2$.

\begin{table}[hbp]
\centering  
\begin{tabular}{cl}  
\hline
Fixed point &Stability condition\\ \hline
$(x_1,y_1)$ &$\lambda<\hat\lambda_{c}^{f},\footnotesize{\alpha \in(\alpha_{0}^{-}, 0)}$\\         
  &$\lambda>0,\footnotesize{\alpha \in(-\pi/2,\alpha_{0}^{-})}$\\
$(x_2,y_2)$ &$\lambda>\lambda_{sc}^{+},\footnotesize{\alpha \in(\alpha_{0}^{+}, \pi/2)}$ \\          
$(x_3,y_3)$ &$\lambda>\lambda_{sc}^{-}, \footnotesize{\alpha \in(\alpha_{0}^{-}, 0)}$ \\
  &$\lambda<\lambda_{sc}^{+}, \footnotesize{\alpha \in(\alpha_{0}^{+}, \pi/2 )}$ \\
$(x_4,y_4)$ & always unstable\\
\hline
\end{tabular}
\caption{The stability conditions of the four fixed points. }
\label{tabel:1}
\end{table}

The fixed points of order parameter equation are related to the collective states of coupled phase oscillators. Fixed points $(x_{3,4},y_{3,4})$ with $|z|=1$ correspond to \textbf{the synchronous state (SS)} of the system where all the phase differences between hub and leaf nodes are the same and keep constant as
\begin{equation}\label{equ:39}
\varphi_{j}(t)=Const, 1\leq j\leq K,
\end{equation}
implying the global synchronization of leaves and the hub in a star network. The above stability analysis indicates that fixed point $(x_{4},y_{4})$ corresponds to the unstable synchronous state and the fixed point $(x_{3},y_{3})$ corresponds to the stable synchronous state, and their stability can be easily studied.

The fixed points $(x_{1,2},y_{1,2})$ with the modulus $|z|=\sqrt{x^{2}+y^{2}}>1$ is unphysical because the order parameter $z$ of coupled oscillators is bounded by $|z|\leq 1$. If $|z|<1$, the related collective state is called \textbf{the splay state (SPS)}~\cite{Strogatz1993,marvel2009invariant}, the phase differences between the hub and leaf nodes satisfy a function relation as
\begin{equation}\label{equ:40}
\varphi_{j}(t)=\varphi(t+\frac{jT}{K}), 1\leq j\leq K
\end{equation}
with $T$ the period of $\varphi(t)$, as shown in Fig. \ref{fig:1}(a).  This kind of state physically represents the collective state where all the leaf oscillators in the star network move synchronously with a constant time shift. 

\section{Collective Dynamics of Time-dependent States}
\label{sec:4}
Long-term solutions of the order parameter equation contains not only the states given by fixed points but also time-dependent states corresponding to periodic solutions. There are two periodic regimes, i.e.,  the regime $0<\alpha<\pi/2$ and $\lambda<\lambda_{ec}=\lambda_2$, and the critical line with $\alpha=0,\pm\pi/2$: 

\textbf{1. The in-phase state}

When $\lambda<\lambda_{ec}$ and $0<\alpha<\pi/2$,  as shown in Fig. \ref{fig:1}(b), the largest Lyapunov exponent is zero and the other exponents are negative, implying a stable limit-cycle solution. This solution can be conveniently found by transforming the Eq. (\ref{equ:30}) to polar coordinates as $z=re^{i\Phi}$:
\begin{equation}\label{equ:41}
\begin{aligned}
&\dot r=-\dfrac{\lambda}{2}(r^2-1)\cos(\Phi+\alpha),\\
&\dot\Phi=-\dfrac{\lambda}{2}(r+\dfrac{1}{r})\sin(\Phi-\alpha)+\Delta\omega-\lambda Kr\sin(\Phi+\alpha).\\
\end{aligned}
\end{equation}
There is a limit cycle solution with $r=1$ and periodic phase $\Phi(t)$, which is called \textbf{the in-phase state(IPS)}, where all the phase differences between leaves and the hub are the same and time dependent, i. e.,
\begin{equation}\label{equ:42}
\varphi_{j}(t)=\varphi(t), 1\leq j\leq K.
\end{equation}
This state corresponds to the phases of all leaf nodes evolve synchronously, while they are not synchronous to the hub, as shown in Fig. \ref{fig:1}(c). In this case, the model Eq. (\ref{equ:23}) is reduced to the case with $K=1$ and the stability of the state can be obtained by Floquet theory for limit cycle as it is stable for $0<\alpha<\pi/2$ and unstable for $-\pi/2<\alpha<0$.
\begin{figure}
  \includegraphics[width=3in]{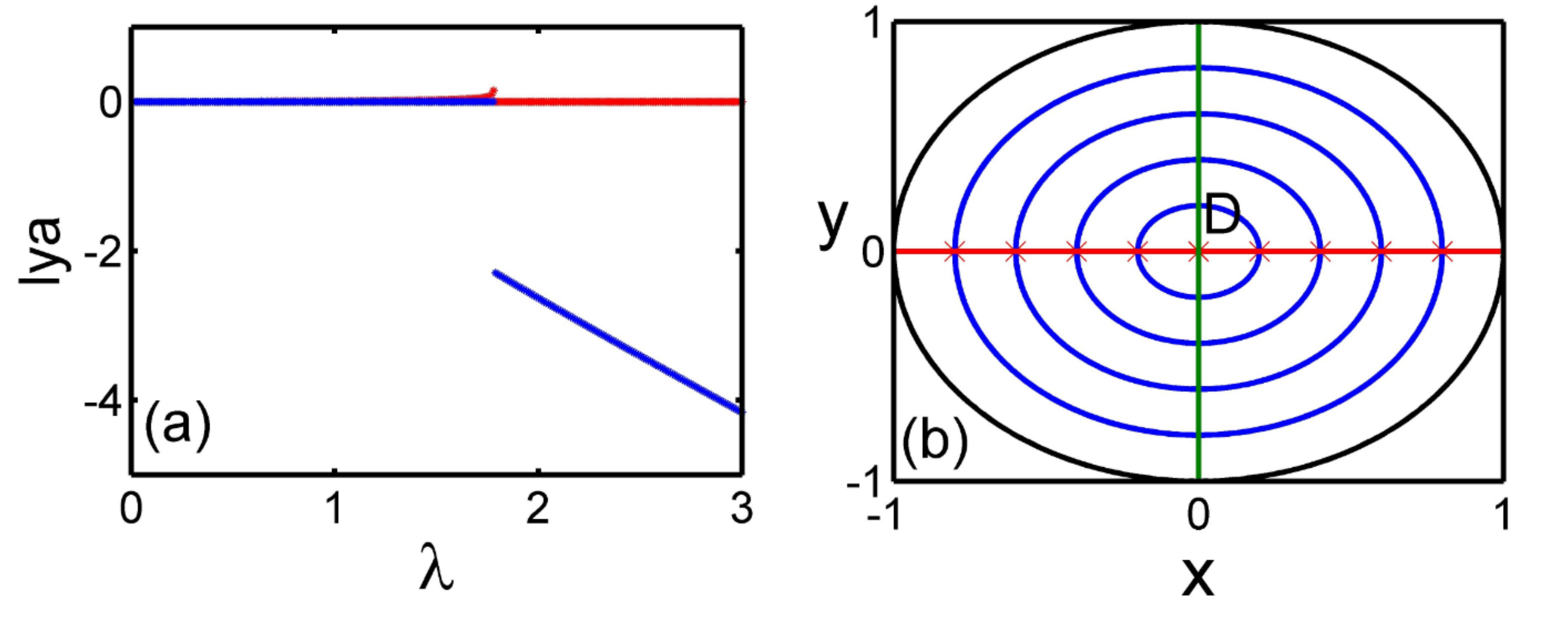}
\caption{($a$) The Lyapunov exponents of the network with $\alpha=0,N=11$. ($b$) Phase plane of Eq. (\ref{equ:31}) with $\Delta\omega=9$, $K=10, \alpha=0$, $\lambda=0.1$. Red lines are $\dot x=0$, and green lines are $\dot y=0$. The intersection of $\dot x=0$ and $\dot y=0$ is fixed point D. Trajectories with different initial values are  marked by '$\ast$'.}
\label{fig:2}       
\end{figure}

\textbf{2.  The neutral state}

The dynamics  at $\alpha=0,\pm\pi/2$ when $\lambda<\lambda_{ec}$ are spacial cases and correspond to the critical dynamical states of the system, where the related fixed point is found to be neutrally stable where Re$(\beta_{1,2})=0$ and Im$(\beta_{1,2})\neq0$ in Eq. (\ref{equ:38}). In this case there are a large class of states in the critical cases with the order parameter $r$ are determined by initial values of $(x,y)$, where long-term behaviors of $z$ depend crucially on initial phases, as shown in Fig. \ref{fig:1}(d). We call this state \textbf{the neutral state(NS)}~\cite{topaj2002reversibility}, and the corresponding fixed point is neutrally stable, where all Lyapunov exponents of the fixed point are zero when $\lambda<\lambda_{ec}$, as shown in Fig. \ref{fig:2}(a). Because the Kuramoto system is dissipative~\cite{florian2012disspetive}, the existence of a large class of neutral states are counterintuitive, which implies that the phase space of this state contains an integrable Hamiltonian system family of periodic orbits, shown in Fig. \ref{fig:2}(b). 

To understand the mechanism of these neutral states, we resort to the analysis of the order parameter equations (\ref{equ:31}). For the case of $\alpha=0$, Eq. (\ref{equ:31}) can be simplified to
\begin{equation}\label{equ:43}
\begin{aligned}
\dot x &=\lambda(K+\frac{1}{2})y^{2}-\frac{\lambda}{2}x^{2}-\Delta\omega y +\frac{\lambda}{2},\\
\dot y &=-\lambda(K+1)xy+\Delta\omega x.
\end{aligned}
\end{equation}
 The fixed points are determined by setting $\dot x=0$ and $\dot y=0$. When $\lambda<\lambda_{ec}$ only the fixed point $(x_1,y_1)$ exists inside the unit cycle in the plane as shown in Fig. \ref{fig:2}(b). The fixed point is neutrally stable, and all the Lyapunov exponents of the the neutral state with $\alpha=0$ are zero. It is worthy to note that if we define a time reversal transformation as $R:(t,x,y)\mapsto(-t,-x,y)$, the dynamical equations (\ref{equ:43}) remain invariant. Hence they are called the time-reversible dynamical system or the quasi-Hamiltonian system~\cite{topaj2002reversibility}. This symmetry endows the system many interesting properties.

Note that, the time reversal transformation $R$ can be resolved into $R=TW$ with $T:t\mapsto-t$ and $W:(x,y)\mapsto(-x,y)$. Hence the invariant set for $W$ is the $y$ axis with $x=0,y>0$. For any trajectory crossing this invariant set, according to the time reversal symmetry, the forward trajectory and the backward trajectory are symmetric. If the forward trajectory evolves to an attractor, the backward trajectory will evolve to the symmetric repeller of the system. Then the attractor and the repeller of the system emerge in pairs. When the trajectory crosses the invariant set more than once, the forward and backward trajectory will coincide with each other, forming the periodic solution for the system, which is called the reversible trajectory~\cite{topaj2002reversibility}. For any reversible trajectory, the Lyapunov exponents have the sign-symmetry form and the volume of phase space in the vicinity of it are conserved in average as we discovered in numerical simulations.

For our order parameter plane of the system, it is bounded by the unit circle with the invariant set as $x=0,y>0$, the attractor and the repeller emerge in the same time, implying that if the plane only exist one fixed point, it is neither the attractor nor the repeller, i. e., the only fixed point is neutrally stable. Suppose that there only exist one neutrally stable fixed point in the plane, the trajectories are vagrant and must cross the invariant set more than once, then those trajectories are closed and periodic. This is what happens when $\alpha=0$ with the region $\lambda<\lambda_{ec}$ as shown in Fig. \ref{fig:2}(b). For $\alpha=0$ and $\lambda>\lambda_{ec}$, there is a coexisting region for the synchronous state and the neutral state as the critical cases for the coexistence region for the neutral state and the synchronous state.

\begin{figure}
 \includegraphics[width=3.5in]{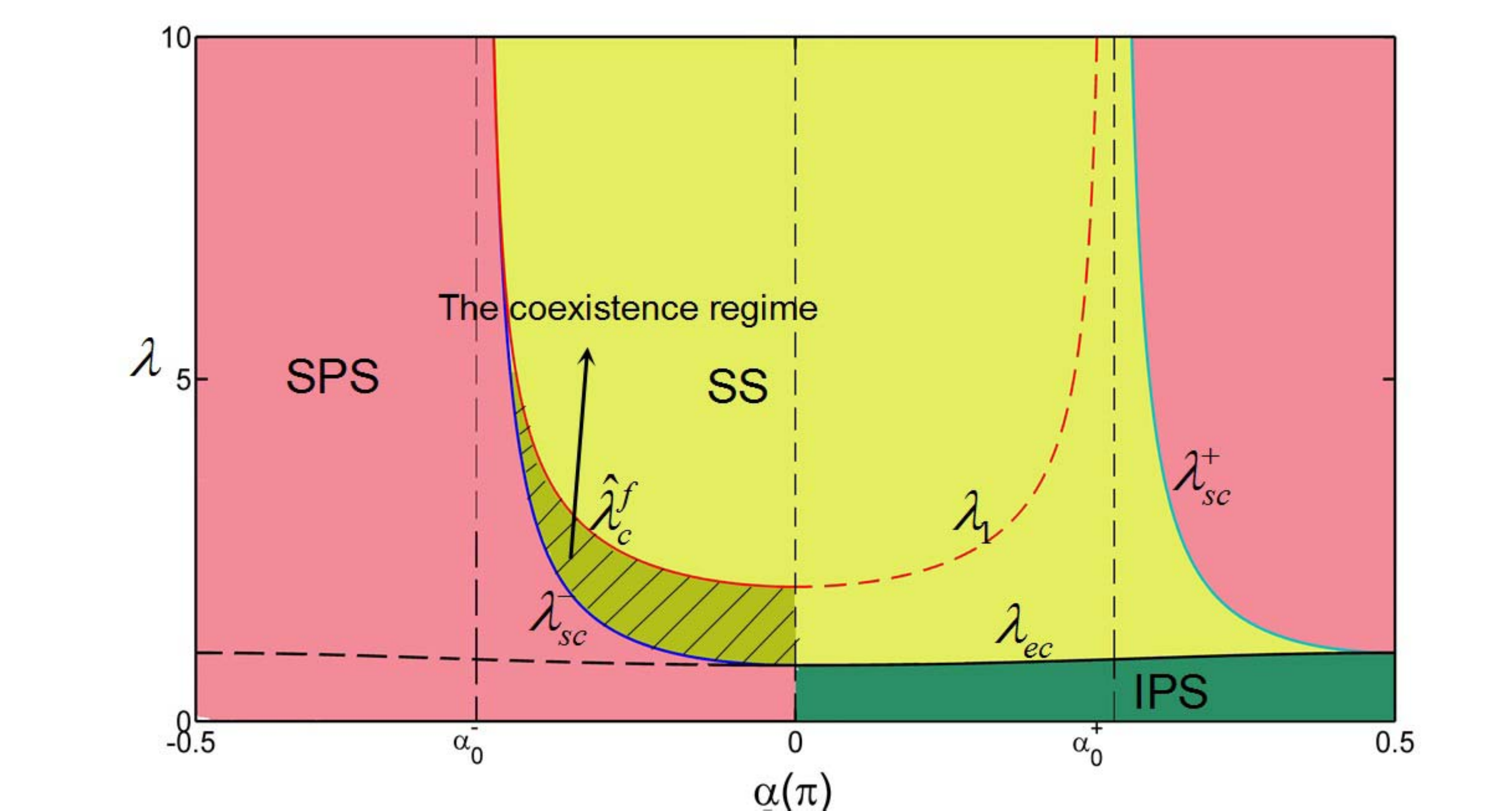}
\caption{Phase diagram of the Sakaguchi-Kuramoto model. Regimes SS, SPS and IPS are stable region for the synchronous state, the splay state and the in-phase state respectively. The stable region for the neutral state is too narrow to plot with only $\alpha=0, \pm\frac{\pi}{2}$. The coexistence regime of the splay state and the synchronous state is plotted by shadow.}
\label{fig:3}       
\end{figure}

All the above possible collective states are summarized in the parameter space ($\alpha$,$\lambda$) as a phase diagram Fig. \ref{fig:3} with the boundaries we get analytically above both from existence and stability conditions. In Fig. \ref{fig:3}, four regions of the phase shift $\alpha$ can be identified. For the first region $-\pi/2<\alpha<\alpha_{0}^{-}$, the splay state exists and is stable for any $\lambda$. With the increase of coupling strength $\lambda$, unstable synchronous state exists above the threshold $\lambda>\lambda_{ec}$. For the second region $\alpha_{0}^{-}<\alpha<0$, the splay state exists and is stable within $0<\lambda<\hat\lambda_{c}^{f}$, and the synchronous state exist with $\lambda>\lambda_{ec}$ but unstable unless $\lambda>\lambda_{sc}^{-}$. Obviously there exist a co-existing region for the splay state and the synchronous state in the region $-\alpha_0^{-}<\alpha<0$, within the coupling interval $\lambda_{sc}^{-}<\lambda<\hat\lambda_{c}^{f}$. In the third region where $0<\alpha<\alpha_{0}^{+}$, the splay state is always unstable, the stable synchronous state emerges as the coupling strength $\lambda>\lambda_{ec}$. For the forth region $\alpha_{0}^{+}<\alpha<\pi/2$, the splay state always exist but only stable when $\lambda>\lambda_{sc}^{+}$, and the synchronous state only exist and stable in the region $\lambda_{ec}<\lambda<\lambda_{sc}^{+}$. The neutral state exists as a particular case for the phase shift, $\alpha=0, \pm\pi/2$, and the in-phase state is always stable in the region $0<\alpha<\pi/2$, within the coupling range $0<\lambda<\lambda_{ec}$. The variety of states in the phase diagram leads to various transitions among them.

\section{Scenarios of Synchronization Transitions}
\label{sec:6}
 The phase diagram shown in Fig. \ref{fig:3} presents a great variety of transitions among the different collective dynamical states. One can also find that some of the states coexist with each other at the same parameter. These coexisting states may lead to abrupt transitions among them and hysteresis behaviors, while the others lead to continuous transitions.

\textbf{1.  Synchronization transition from the neutral state}

\begin{figure}
 \includegraphics[width=3in]{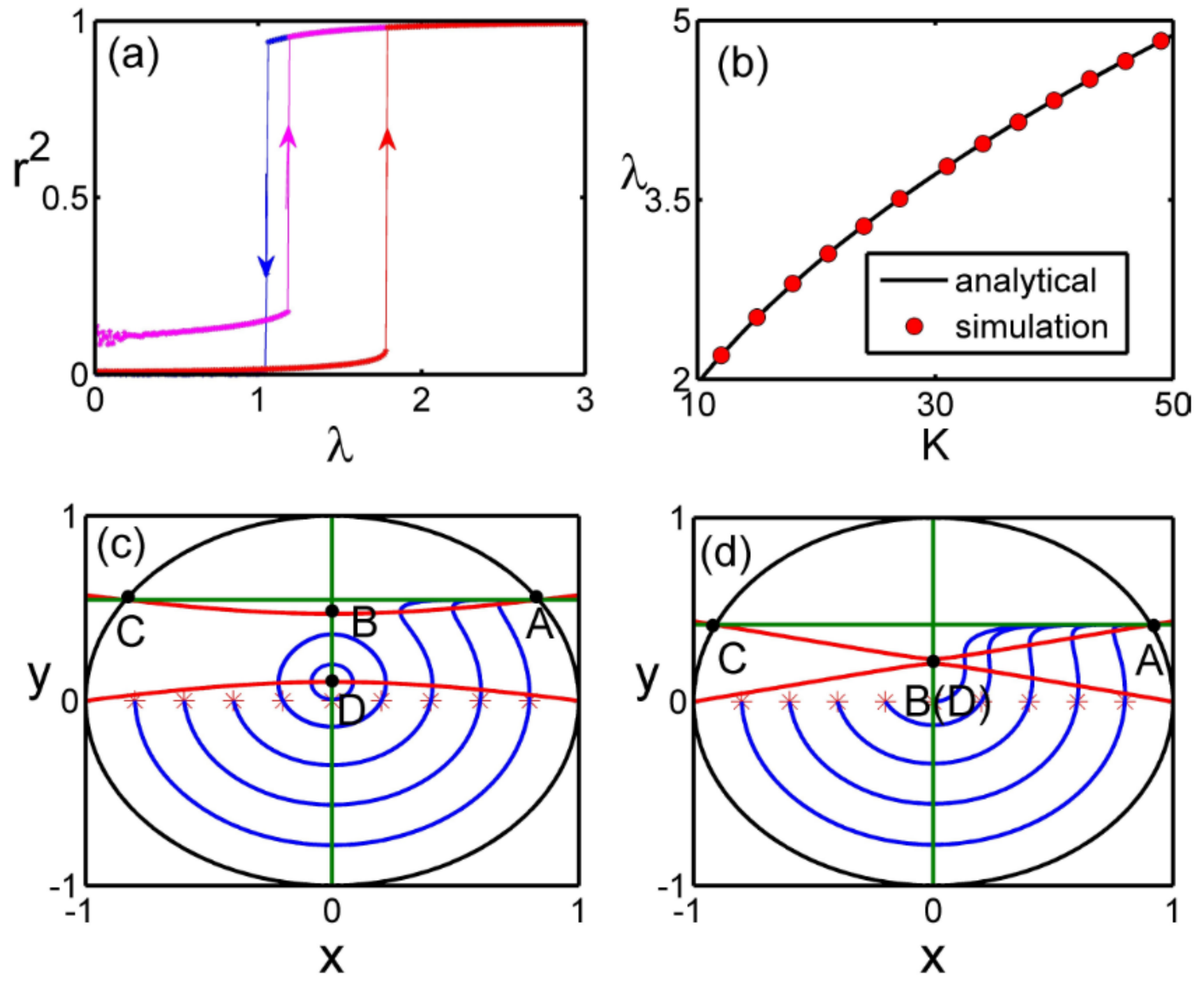}
\caption{($a$) The forward and backward continuation diagrams with $\alpha=0, N=11$. ($b$) The upper limit of forward critical coupling strength with $\alpha=0$ in Eq. (\ref{equ:44}). Phase plane of Eq. (\ref{equ:31}) with $\Delta\omega=9$, $K=10, \alpha=0$, ($c$) $\lambda=1.5$, ($d$) $\lambda=1.9$. Red lines are $\dot x=0$, and green lines are $\dot y=0$. The intersections of $\dot x=0$ and $\dot y=0$ are fixed points A, B, C, D. Trajectories with different initial values are  marked by '$\ast$'.}
\label{fig:4}       
\end{figure}

We first investigate the synchronization process from the neutral state to the synchronous state for $\alpha=0$. The synchronization process when $\alpha=0$ is discontinuous known as the explosive synchronization which has attracted much attention recently~\cite{gomez2011explosive}. It has been shown that with changing the coupling strength $\lambda$, this kind of transition is abrupt, and there is a hysteretic behavior at the onset of synchronization, and $\lambda_{c}^{b}$ and $\lambda_{c}^{f}$ are the backward and forward critical coupling strengths respectively, where $\lambda_{c}^{b}=\lambda_{2}$ and  $\lambda_{c}^{f}$ depends on initial states as shown in Fig. \ref{fig:4}(a). The upper limit of $\lambda_{c}^{f}$ is denoted by $\hat\lambda_{c}^{f}$. As $\lambda>\hat\lambda_{c}^{f}$, the synchronization state is globally attractive. It is difficult to understand this process on the basis of the self consistent method, especially for the hysteresis behavior and coexisting region.

The critical coupling corresponds to the upper limit of $\lambda_{c}^{f}$, which can be determined as
\begin{equation}\label{equ:44}
\hat{\lambda_{c}^{f}}=\frac{\Delta{\omega}}{\sqrt{2K+1}}.
\end{equation}
The analytical curve and the simulation results are given in Fig. \ref{fig:4}(b), it is clear that the results conform with the curve.

In the bistable regime, as shown in Fig. \ref{fig:4}(c), the nullclines $\dot{x}=0$ (the red lines) and $\dot{y}=0$ (the green lines) have four intersections labeled by A-D with A an attractor, C a repeller and B, D neurally stable. Any orbits crossing the nullcline A-B-C will eventually fall to A, and others will hold the property as periodic orbits. It is clear that the stable fixed point A corresponds to the synchronous state. And the basin for the neutral state can be calculated approximated by the circle which has its center in point $D$ and radius as the length of line $B-D$. As $\lambda$ increases, points D and B close to each other and eventually collide at a critical coupling, as shown in Fig. \ref{fig:4}(d), and the synchronous state becomes globally attractive. 

\textbf{2.  Synchronization transition from the splay state}

\begin{figure}
 \includegraphics[width=3in]{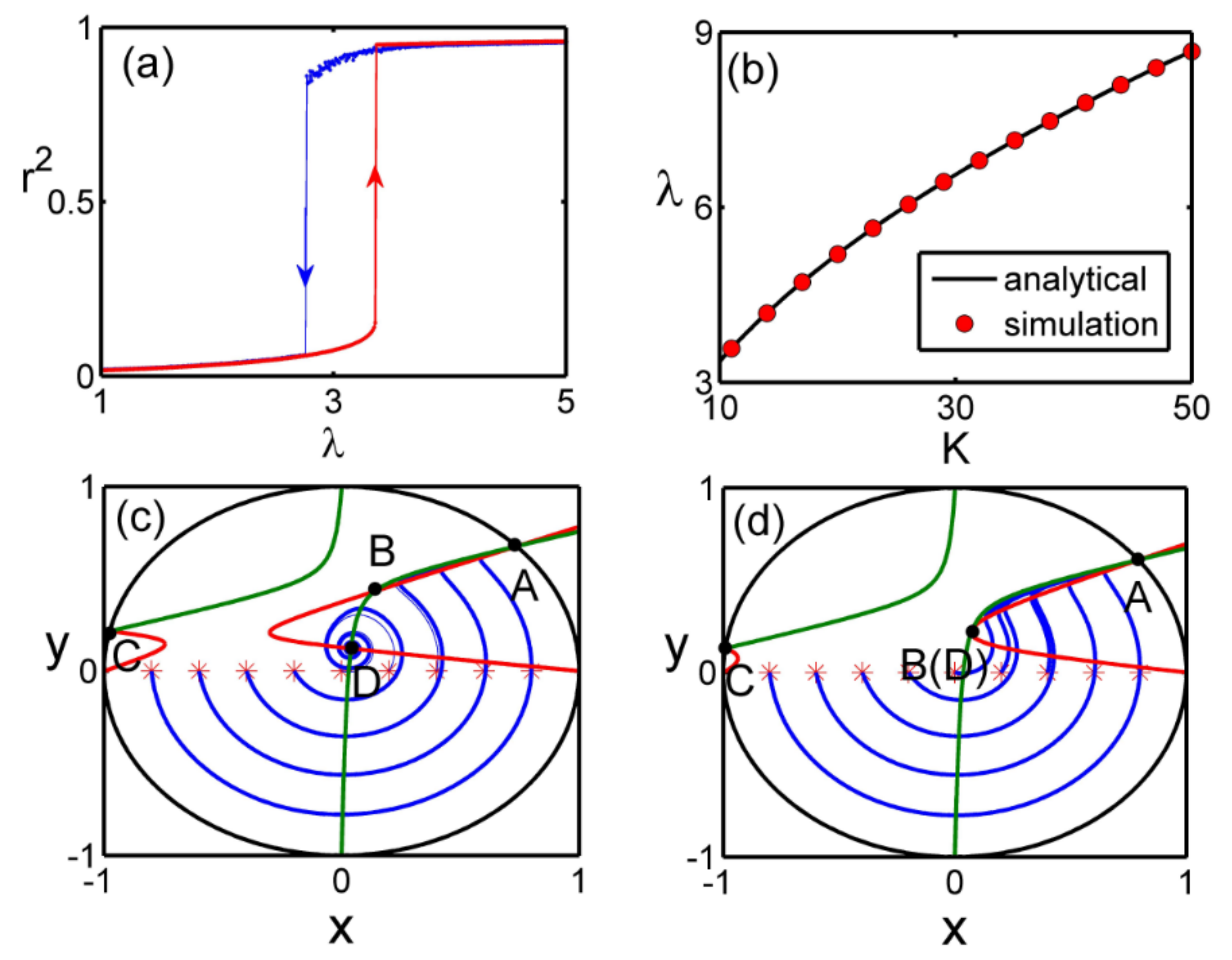}
\caption{($a$) The forward and backward continuation diagrams with $\alpha=-0.2\pi, N=11$. ($b$) The upper limit of forward critical coupling strength with $\alpha=-0.2\pi$ in Eq. (\ref{equ:45}). Phase plane for $\Delta\omega=9$, $K=10$, $\alpha=-0.1\pi$, ($c$) $\lambda=1.8$, ($d$) $\lambda=2.17$. Red lines are $\dot x=0$ and green lines are $\dot y=0$. The intersections of $\dot x=0$ and $\dot y=0$ are the fixed points A,B,C,D. Trajectories with different initial values are marked as '$\ast$'.}
\label{fig:5}       
\end{figure}

The synchronization process from the splay state to the synchronous state for $\alpha_{0}^{-}<\alpha<0$ is found to be discontinuous. Numerical computations reveal that this kind of transition is abrupt with hysteresis at the onset of synchronization as shown in Fig. \ref{fig:5}(a). The abrupt transition implies that there are two critical coupling strengths $\lambda_{c}^{b}$ and $\lambda_{c}^{f}$, where $\lambda_{c}^{b}=\lambda_{sc}^{-}$ and $\lambda_{c}^{f}$ depend on the basin of attraction. The upper limit of $\lambda_{c}^{f}$ can be determined by analyzing the inverse saddle-node bifurcation as
 \begin{equation}\label{equ:45}
\hat{\lambda_{c}^{f}}=\frac{\Delta{\omega}}{\sqrt{2Kcos(2\alpha)+1}}.
\end{equation}
As shown in Fig. \ref{fig:5}(b), the simulation results are consistent with the analytical curve.

The dynamical manifestations of the discontinuous transition from the splay state to the synchronous state are shown in Fig. \ref{fig:5}(c,d). Fig. \ref{fig:5}(c) exhibits the coexistence of the splay state and the synchronous state as the stable fixed points D and A respectively. The basins of attraction of the splay state and the synchronous state are separated by the saddle point B. When coupling $\lambda$ increases, as shown in Fig. \ref{fig:5}(d), the saddle point B and the attractor D collide and disappear via an inverse saddle-node bifurcation, and this discontinuous transition makes the fixed point A corresponds to the synchronous state a global attractor.

\textbf{3.  Synchronization transition from the in-phase state}

\begin{figure}
 \includegraphics[width=3in]{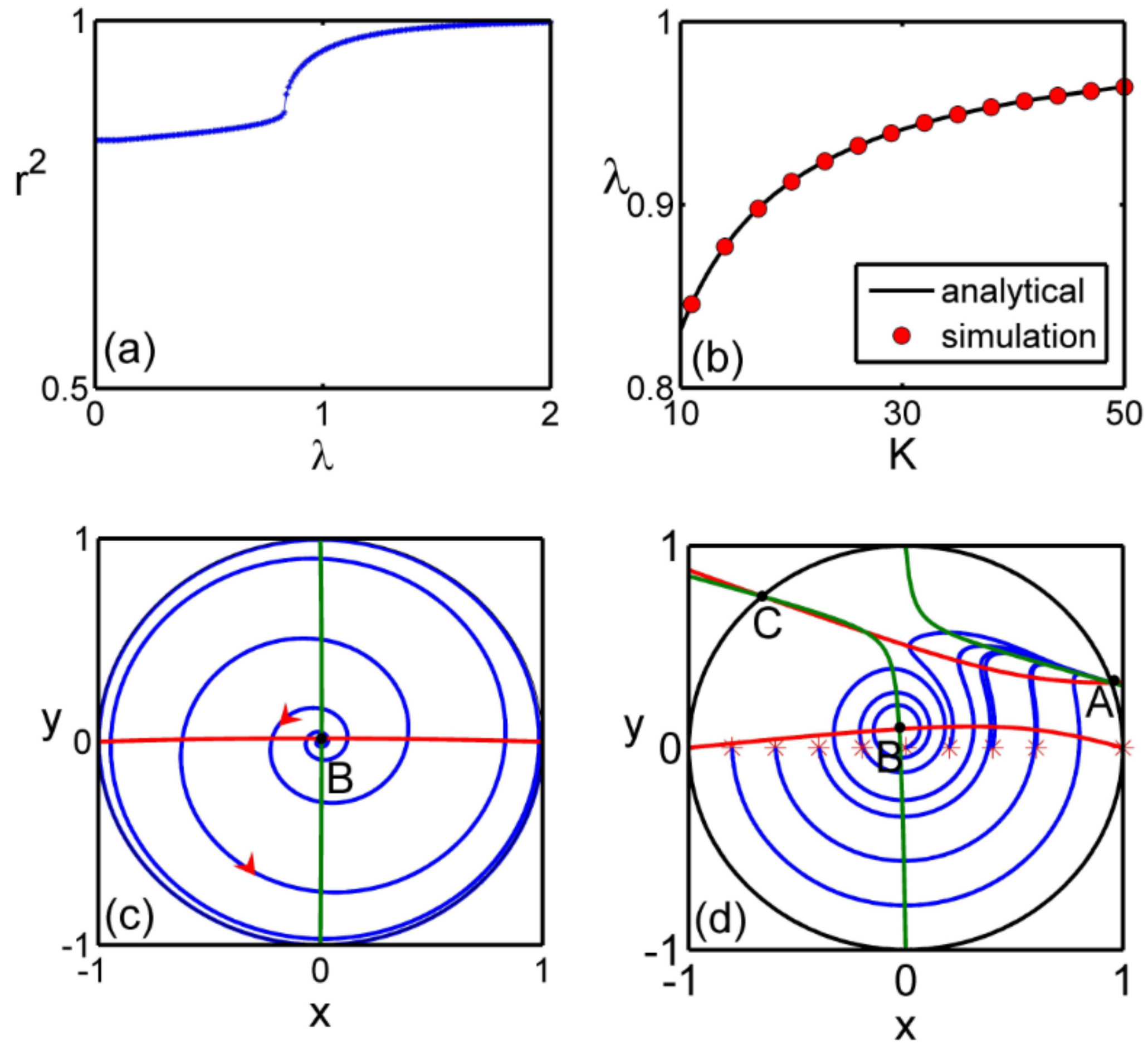}
\caption{($a$) The forward continuation diagrams with $\alpha=0.3\pi, N=11$. ($b$) The forward critical coupling strength with $\alpha=0.3\pi$ in Eq. (\ref{equ:35}). Phase plane for $\Delta\omega=9$, $K=10$, $\alpha=0.3\pi$, ($c$) $\lambda=0.5$, ($d$) $\lambda=1.5$. Red lines are $\dot x=0$ and green lines are $\dot y=0$. The intersections of $\dot x=0$ and $\dot y=0$ are the fixed points A,B,C,D. Trajectories with different initial values are marked as '$\ast$'.}
\label{fig:6}       
\end{figure}

The route of synchronization from the in-phase state to the synchronous state for $\alpha>0$ is shown in Fig. \ref{fig:6}(a).  The critical coupling strength of this continuous transition $\lambda_{ec}$ is determined by Eq. (\ref{equ:35}). It can be found from Fig. \ref{fig:6}(b) that the simulation results agree well with the analytical curve. 

The dynamical manifestations of the transition from the in-phase state to synchronous state are shown in Fig. \ref{fig:6}(c,d). As shown in Fig. \ref{fig:6}(c), the in-phase state is a limit cycle in the order parameter plane. As $\lambda$ increases, the stable fixed point A corresponding to the synchronous state emerges on the limit cycle. The transition from the in-phase state to the synchronous state takes place continuously through a saddle-node bifurcation, as shown in Fig. \ref{fig:6}(d).

\textbf{4.  Scenario of desynchronization}

\begin{figure}
 \includegraphics[width=3in]{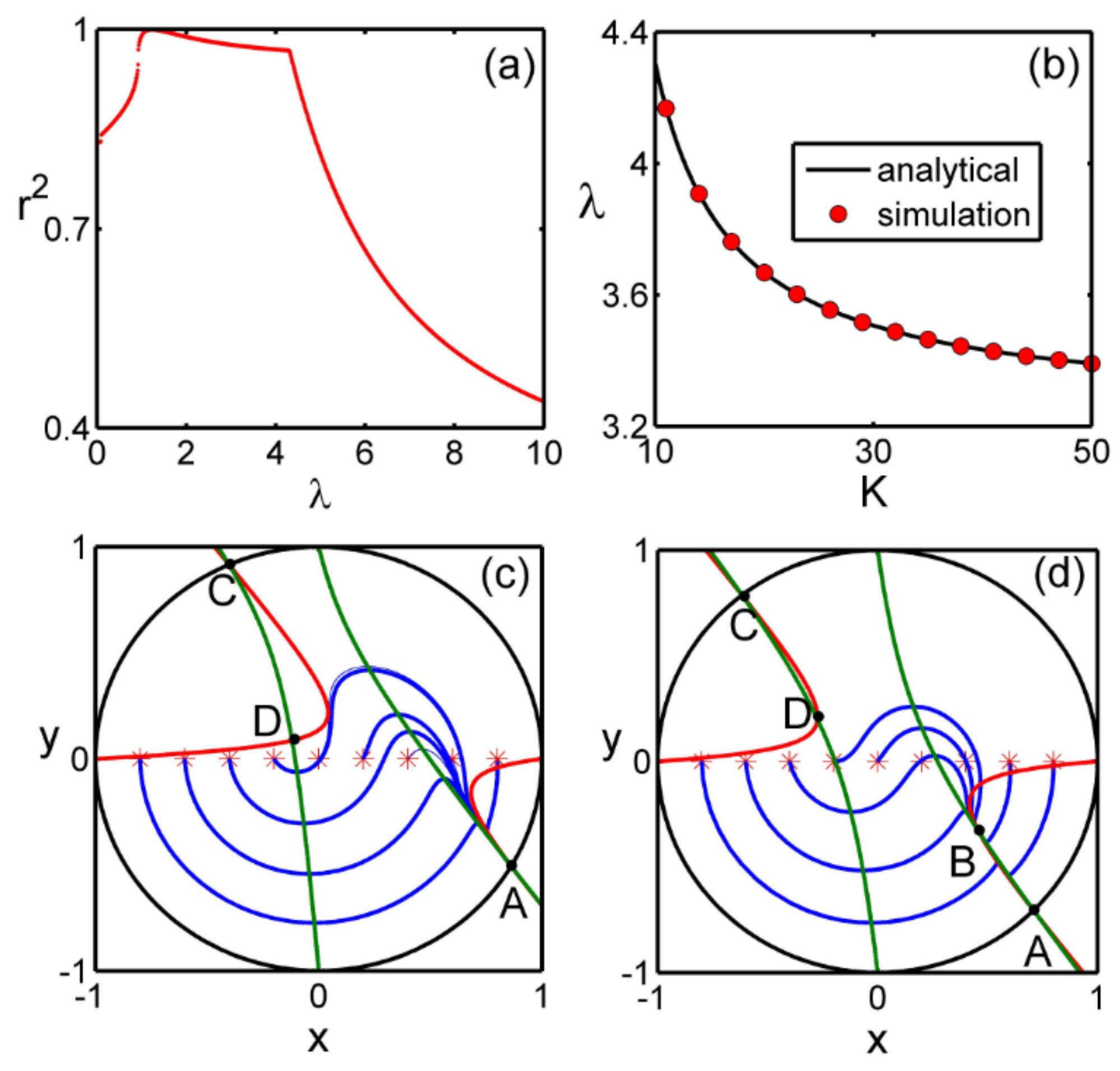}
\caption{($a$) The order parameter against the coupling strength with $\alpha=0.3\pi, N=11$. ($b$) The critical coupling strength $\lambda_{sc}^{+}$ with $\alpha=0.3\pi$. Phase plane for $\Delta\omega=9$, $K=10$, $\alpha=0.3\pi$, ($c$) $\lambda=3$, ($d$) $\lambda=5$. Red lines are $\dot x=0$ and green lines are $\dot y=0$. The intersections of $\dot x=0$ and $\dot y=0$ are the fixed points A,B,C,D. Trajectories with different initial values are marked as '$\ast$'.}
\label{fig:7}       
\end{figure}

In the region $\alpha_{0}^{+}<\alpha<\frac{\pi}{2}$ of the phase diagram \ref{fig:3}, one may find the synchronous state is unstable when $\lambda>\lambda_{sc}^{+}$ and the stable splay state emerges, which is contrary to our conventional belief that the system will always be synchronous if the coupling strength is large enough.
 The transition is called the desynchronization, and it is a continuous transition as shown in Fig. \ref{fig:7}(a).
 The order parameter $r$ decrease rapidly at the threshold and effective frequencies of hub and leaf nodes are divided at the same coupling $\lambda$. It is easy to know the route of the de-synchronization is from the synchronous state to the splay state from the view of the phase diagram. The threshold of the de-synchronization is $\lambda_{sc}^+=\footnotesize{{-\Delta\omega}/({K\cos2\alpha+1}})$ as shown in Fig. \ref{fig:7}(b), the simulation results are consistent with it obviously. 

The dynamical manifestations of the continuous transition from the synchronous state to the splay state are shown in Fig. \ref{fig:7}(c,d). Fig. \ref{fig:7}(c) exhibits the stable synchronous state of the system when $\lambda<\lambda_{sc}^{+}$, all the orbits in the phase of the order parameter will evolve to the fixed point A eventually. As $\lambda$ increases and larger than the critical coupling $\lambda_{sc}^{+}$, the two nullclines will intersect in four fixed points as shown in Fig. \ref{fig:7}(d), the point A loses its stability and a new stable fixed point B which corresponds to the splay state appears. The process from the synchronous state to the splay state is finished by this bifurcation continuously.

\section{Conclusion}

To summarize, in this paper we study the dynamics of coupled oscillators on a star network with the Sakaguchi-Kuramoto model by resorting to the dynamical order parameter equation that can be obtained in terms of different approaches, e.g.,the ensemble order parameter approach and the Watanabe-Strogatz approach. The order parameter equation obtained for star network can also be approximately described from the Ott-Antonsen ansatz, which is originated from the high symmetry of the topology. By reducing from a high-dimensional phase space to a much lower-dimensional order parameter space without additional approximation, one is able to grasp analytically the essential dynamical mechanism of different scenarios of synchronization. Different solutions of the order parameter equation corresponds to the various collective states of coupled oscillators, and different bifurcations reveal various transitions among those collective states. The process of those transitions are revealed in the plane of order parameter and the critical coupling strengths of them are obtained analytical which are verified by the simulation results.

This work is partially supported by the National Natural Science Foundation of China (Grant No. 11075016 and 11475022) and the Scientific Research Funds of Huaqiao University.

\end{document}